\documentclass{article}

\usepackage{arxiv}

\usepackage[utf8]{inputenc} 
\usepackage[T1]{fontenc}    
\usepackage{hyperref}       
\usepackage{url}            
\usepackage{booktabs}      
\usepackage{amsfonts}       
\usepackage{nicefrac}       
\usepackage{microtype}      
\usepackage{lipsum}		
\usepackage{graphicx}
\usepackage{doi}
\usepackage{amssymb}
\usepackage{amsmath}
\usepackage{url}
\usepackage{eurosym}
\usepackage{tabularx}
\usepackage{array}
\usepackage{multicol}
\usepackage{float}
\usepackage{caption}
\usepackage[backend=biber, style=authoryear]{biblatex}
\newcolumntype{C}[1]{>{\centering\arraybackslash}p{#1}}

\title{An Online Hierarchical Energy Management System for Energy Communities, Complying with \\ the Current Technical Legislation Framework}

\author{ \href{https://orcid.org/0000-0002-6360-7737}{\includegraphics[scale=0.06]{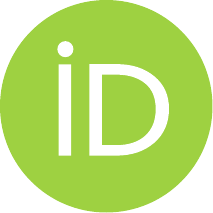}\hspace{1mm}Antonino~Capillo}\\
	DIET Dept.\\
	University of Rome ``La Sapienza”\\
	Rome, IT, Via Eudossiana 18, 00184 \\
	\texttt{antonino.capillo@uniroma1.it} \\
	\And
	\href{https://orcid.org/0000-0003-4915-0723}{\includegraphics[scale=0.06]{orcid.pdf}\hspace{1mm}Enrico ~De Santis} \\
	DIET Dept.\\
	University of Rome ``La Sapienza”\\
	Rome, IT, Via Eudossiana 18, 00184 \\
	\texttt{enrico.desantis@uniroma1.it} \\
 	\And
	\href{https://orcid.org/0000-0002-3748-5019}{\includegraphics[scale=0.06]{orcid.pdf}\hspace{1mm}Fabio M. ~Frattale Mascioli} \\
	DIET Dept.\\
	University of Rome ``La Sapienza”\\
	Rome, IT, Via Eudossiana 18, 00184 \\
	\texttt{fabiomassimo.frattalemascioli@uniroma1.it} \\
 	\And
	\href{https://orcid.org/0000-0001-8244-0015}{\includegraphics[scale=0.06]{orcid.pdf}\hspace{1mm}Antonello ~Rizzi} \\
	DIET Dept.\\
	University of Rome ``La Sapienza”\\
	Rome, IT, Via Eudossiana 18, 00184 \\
	\texttt{antonello.rizzi@uniroma1.it} \\
}

\renewcommand{\shorttitle}{\textit{arXiv} Template}

\hypersetup{
pdftitle={A template for the arxiv style},
pdfsubject={cs.AI, cs.AI},
pdfauthor={Antonino~Capillo, Enrico~De Santis, Fabio M. ~Frattale Mascioli, Antonello ~Rizzi},
pdfkeywords={energy management system, smart grid, renewable energy community, evolutionary optimization, fuzzy inference system, ai explainability},
}

\begin{document}
\maketitle

\begin{abstract}
Efforts in the fight against Climate Change are increasingly oriented towards new energy efficiency strategies in Smart Grids (SGs). In 2018, with a proper legislation, the European Union (EU) defined the Renewable Energy Community (REC) as a local electrical grid whose participants share their self-produced renewable energy aiming at reducing bill costs by taking advantage of proper incentives. That action aspires to be an accelerator to the spread of local renewable energy exploitation, whose costs could not be within everyone's reach. Since a REC is technically a SG, the aforementioned strategies can be applied and, specifically, effective Energy Management Systems (EMSs) are required. Therefore, in this work, an online Hierarchical Energy Management System (HEMS) is synthesized for REC cost minimization to evaluate its superiority over a local self-consumption approach. EU technical indications (as inherited from Italy) are diligently followed aiming at as realistic as possible results. Power flows between REC nodes or Microgrids (MGs) are optimized by taking Energy Storage Systems (ESSs) and PV plant costs, energy purchase costs and REC incentives. A hybrid Fuzzy Inference System - Genetic Algorithm (FIS-GA) model, with the GA encoding the FIS parameters, is implemented. Power generation and consumption, which are the overall system input, are predicted by a Long Short-Term Memory Neural Network (LSTM), trained on historical data. The proposed hierarchical model achieves good precision in short computation times and it outperforms the self-consumption approach leading to about $20 \thinspace \%$ savings compared to the latter. In addition, the Explainable AI (XAI), which characterizes the model through the FIS, makes results more reliable thanks to a good human interpretation level. To finish, the HEMS is parametrized so that it is very simple to switch to another Country's technical legislation framework that, together with its good precision and computation time, should make it worthy of further development also by considering a proper business model for real-life applications.
\end{abstract}


\section{Introduction}
\label{sec:introduction}

\noindent Among the proper countermoves against Climate Change, energy efficiency strategies in Smart Grids (SGs) are relevant. In 2018, with the RED II ~\cite{red_II} directive, the European Union (EU) defined the Renewable Energy Community (REC) as a local electrical grid whose participants share their self-produced renewable energy aiming at reducing bill costs by taking advantage from specific incentives. That aspires to be very useful for making Renewable Energy Source (RES) technologies more affordable for everyone by helping to face often too much expensive RES generators and Energy Storage System (ESS) installation and wear. Indeed, being grouped together, REC participants could save on electricity bills. However, without a suitable EMS in charge of making optimal decisions about energy flows (e.g. about when and how much to charge the ESS), the above savings, if they occurred, might not be appreciable. Being RECs a subset of SGs in all respects, both because of its architecture and its operating principle, the EMSs technology is widely applicable to them. \\

\noindent EMSs in Smart Grids (SGs) are extensively discussed in the recent literature. Cost minimization in terms of SGs elements operational wear is a very important objective that real-time EMSs ~\cite{a_two_layer} have to achieve in a short time. In fact, online optimization procedures should be followed to avoid forecast errors due to long prediction horizons. In this respect, Neural models are often adopted for power predictions (e.g. load profiles) in support of EMSs ~\cite{a_new_framework}~\cite{a_novel_reinforced}. Likewise, Machine Learning (ML) comes to help in the optimal decision processes ~\cite{safe_deep}, to face problem complexity at the cost of a reasonable approximation.

As regards EMSs in Energy Communities (EC), even though it is a relatively new research topic, a not negligible volume of works can be found in literature.

One of the most common EMS objectives is cost minimization ~\cite{a_three_stage}~\cite{quantifying_the_benefits}, i.e. the energy purchase cost from the Main Grid, PV plant and ESS installation costs and also operational costs like, for example, the ESS wear cost as expressed in ~\cite{predictive_control_and}. In ~\cite{day_ahead_scheduling}, energy selling is taken into account in the problem formulation and that helps to amortize the expenditure. However, the REC self-consumption, which logically speaking distances itself from an energy trading with the Main Grid ~\cite{day_ahead_scheduling}, is often the energy management operation purpose. As an example, in ~\cite{optimal_scheduling_for}, energy flows between the overall ECs and the Main Grid are minimized and in ~\cite{optimizing_planning_and}, similarly, excess in PV energy production is reduced as much as possible. It is not unusual to find some works that contemplate also $CO_2$ emissions minimization as an objective ~\cite{optimal_dispatch}~\cite{mixed_integer_linear}, generally in addition to the above aims, in a multi-objective fashion.

For the sake of realism, technical legislation framework dictates can be included in the problem formulation ~\cite{mixed_integer_linear}, although only in a fraction of the reviewed works about ECs EMSs. In ~\cite{predictive_control_and} there is not a strong reference to EC regulations and some mention to incentives is made in ~\cite{design_and_optimal}. Moreover, in ~\cite{a_three_stage}, an EC business model is investigated in detail by involving the roles of Community Manager and Distribution System Operator (DSO).

Like a SG, an EC is composed of nodes or Migrocrids (MGs), that are the participants or \textit{prosumers} in the incentive mechanism. Besides the fundamental EC elements, (loads - the participants or \textit{prosumers} - RESs generators and ESSs, with a possible connection to the Main Grid), Electric Vehicles (EVs) can be added for a Vehicle-to-Grid (V2G) strategy ~\cite{risk_based}~\cite{coalitional_game}~\cite{co_optimization}. Moreover, architectures equipped with fuel-cell generators and micro-CHP ~\cite{smart_energy_community} can also be found in the literature. Focusing on the ESS, amid the other elements, is essential for a realistic representation of the ECs operations, since it represents the energy buffer in a RES context. Both a centralized ESS and a distributed ESS (one ESS for each prosumer) strategies are being explored. About the former, in ~\cite{a_two_stage_igdt} a centralized ESS consists in a whole EVs parking area, where it is possible to exploit groups of vehicles as batteries (V2G). In ~\cite{predictive_control_and} a distributed strategy is adopted and a wear cost model for the ESS is implemented. More precisely, the aforesaid cost is directly proportional to the ESS exchanged power by a wear cost coefficient. However, an ESS wear cost model is not always implemented, as in ~\cite{a_novel_distributed}.

The strategy scope, i.e. the particular task the EMS simulation is conducted to accomplish, can be both large and slight. In the former case, optimal energy flows between nodes or, at most, between them and the Main Grid, are settled ~\cite{optimal_dispatch}. In the latter case, a more detailed optimal energy flow network is decided, fulfilling the Demand-Response task by shifting controllable household appliances ~\cite{flexibility}~\cite{coalitional_game}~\cite{an_efficient_energy}. Furthermore, for what concerns the simulation time horizon, real-time ~\cite{predictive_control_and} and day-ahead ~\cite{day_ahead_scheduling} EMSs for ECs are both provided in the literature. 

As concerns the technical side, many EMSs for ECs rely on Mixed-Integer Linear Programming (MILP) ~\cite{flexibility}~\cite{local_energy}~\cite{optimal_dispatch} and Alternating Direction Method of Multipliers (ADMM) ~\cite{a_three_stage}~\cite{day_ahead_scheduling} as optimization paradigms. However, Computational Intelligence (CI) is confirmed to be very useful for facing problem complexities and OF non differentiability. In fact,  Genetic Algorithms (GAs), multi-objective purpose Non-dominated Sorting Genetic Algorithms (NSGA-II and NSGA-III) ~\cite{optimizing_planning_and}~\cite{an_efficient_energy}~\cite{co_optimization}~\cite{a_novel_distributed} and Monte Carlo Tree algorithms ~\cite{design_and_optimal} are implemented in a not negligible number of works on the topic at hand.

Also, hierarchical energy management is successfully adopted in the field. In ~\cite{local_energy}, a MILP local EMS (i.e. one for each MG) aimed at costs minimization returns household appliance profiles that are passed as an input to a global EMS (MILP-based) whose objective is the whole EC self-consumption. In ~\cite{two-stages}, a REC cost minimization by an energy management strategy is performed in France. Both global REC-level and single-participant bill savings are considered by making many attempts not only to achieve a good global REC cost minimization but also good electricity bill savings for each participant, encouraging future participants to join. Such a two-level optimization is possible only through a realistic business model that provides details about the REC environment (i.e. relations between REC Manager, participants and suppliers, community payments and billing information, and so on).\\

\noindent In the Italian context, the ESS is confirmed to be a crucial element ~\cite{business_models}~\cite{pantelleria}~\cite{reinforcement_rec}. In ~\cite{reinforcement_rec}, the decision variable is a real-valued number that rules the amount of energy to exchange with the ESS. In ~\cite{business_models}, a power flows optimization, based on the optimal ESS management, is operated for a hypothetical REC in Italy, also by taking into account different business models. Nevertheless, ESS wear modeling is often not treated in deep. In ~\cite{italian_framework}, no explicit ESS operational cost model is implemented but ESS OPEX are assumed to be a function of capital expenditure. Also in ~\cite{business_models}, the ESS OPEX is assumed to be a fixed quantity without a wear model behind it. In ~\cite{pantelleria}, a proper wear battery cost model is not included in the problem formulation since the ESS can perform one charge/discharge cycle per day and a fixed number of maximum lifespan cycles number is set. The same can be said for ~\cite{reinforcement_rec} and ~\cite{system_incentives}, where a battery cost is included into he formulation although a wear model is not directly explicated.

In line with many aforementioned abroad works the EMS objective is often the minimization of energy exchanges with the Main Grid ~\cite{ga_REC}~\cite{power_sharing_model} but in ~\cite{italian_context} the overall REC costs are minimized, like in ~\cite{business_models} and in ~\cite{pantelleria} where an overall REC cost minimization is performed for a site in the Pantelleria island. However, in ~\cite{planning_of_a}, electricity bill costs minimization is pursued by self-consumption maximization.

For what concerns simulation realism and reference to the EC legislation, one of the most detailed works is ~\cite{italian_framework}, where high care is taken in embedding the Italian legislation framework in the problem formulation. Nonetheless, the energy selling price to the Main Grid is only estimated as the mean hourly Italian Market value using 2019 time series, rather than taking it from the technical legislation. That is reasonably due to the legislation updates that occurred from 2020 (when the aforementioned work was written) to the present day. In addition, PV production data are estimated using PVGIS tool, (a European Commission empirical model that takes geographical coordinates as an input) in place of measured data. Since that work mainly aims at the REC PV and ESS sizing, rather than energy management as such, a proper power forecasting model is not implemented.

As for the technical sphere, both exact and heuristic optimization paradigms are adopted. In ~\cite{italian_framework} a MILP optimization algorithm is implemented. Similarly, in ~\cite{italian_context} an exact optimization model is developed in the MathWorks\textsuperscript{\textregistered} Matlab environment. ML and CI are preferred in ~\cite{reinforcement_rec}, where a Reinforcement Learning (RL) EMS is synthesized, and in ~\cite{ga_REC}, where a GA-based EMS, developed in Python, is in charge of optimally control in a REC with also Electric Vehicles (EVs). 

Speaking of nodes, architecture both distributed (PV roofs distributed among participants) and centralized (one PV plant for the whole Community) configurations are explored ~\cite{intalian_context}, and for optimal power flow decisions a real-time approach is generally preferred to the day-ahead scheduling, as in ~\cite{ga_REC}.

\noindent Being ML and CI (so, in general, Artificial Intelligence - AI - more and more used in SGs and ECs EMSs), as explained above, it is worth mentioning the Explainable Artificial Intelligence (XAI) research field ~\cite{critical_thinking}~\cite{explainable_ai_for}~\cite{a_review_of}~\cite{harnessing_explainable}. Instead of relying on black-box models (e.g. Neural Networks), even though their efficiency is undisputed, a grey-box AI algorithm could bring more information about its inner reasoning process. In this context, Fuzzy Logic (FL) is very useful thanks to the structure of Fuzzy Inference Systems (FISs) represented in a way that is very close to Natural Language ~\cite{constrained_interval}. \\

\noindent To sum up, to the best of our knowledge, the following considerations can be made. The real-time approach is the best choice for more realistic results and the alternation of self-consumption and cost minimization as REC global objectives is a constant also in the Italian context. The same is true for the technical side, with ML and CI sometimes preferred to exact optimization (much less in Italy) but there is no mention of XAI in the topic at hand (EMSs in ECs). Moreover, ESS wear model implementation can not be found easily in Italy and it is rarely done in depth abroad. As for legislation references, many efforts have been made with remarkable results but it is difficult to find a completely explicated and updated technical legislation reference in the literature. To finish, the hierarchical energy management research line is still not developed in Italy, at least for the topic at hand.\\

\noindent In this work, an online Hierarchical Energy Management System (HEMS) is synthesized for a REC cost minimization. Each participant is equipped with a local EMS that maximized its self-consumption while the global HEMS, aware of the nodes' state, overwrites the local optimal decisions about power flows. That design choice comes from the need to explore more in deep which objective could be more proper for a REC. None of that could be possible without the realism level given by the slavish reference to the EU technical legislation framework, (as inherited by Italy) as well as an ESS wear cost model more complex than the others in the ECs EMSs literature for Italy, at the best of our knowledge. This paper does not include a REC business model. Therefore, revenues and costs are calculated realistically but only for having a yardstick for a comparison between the self-consumption and the hierarchical strategies. It means that economic details are not covered so that final electricity bill costs are not evaluated. In line with the literature, the HEMS relies on ML both for PV power predictions, performed by an LSTM, and for optimization that uses the FIS-GA paradigm in a XAI fashion. More precisely, the GA implemented in this work comes from a development path followed by the authors that led to previous papers about text classification ~\cite{de_santis_granulation} and also energy management tasks ~\cite{capillo_e_boat}. Since that was a previous version of the algorithm, the updated GA used in this work is accurately re-tested as a stand-alone algorithm on some of the best-known benchmark OFs. Moreover, the use of Fuzzy Logic as an Explainable AI paradigm (especially hybridized with Evolutionary Optimization - EO -) is part of the research team background ~\cite{de_santis_fis}. \\
The main contributions of this paper are the following:

\begin{itemize}
    \item Local self-consumption outperformed by the proposed hierarchical management strategy ($20 \thinspace \%$ savings more);
    \item Slavish reference to the up-to-date EU REC Legislation framework, as inherited by Italy;
    \item Proper model parametrization in order to an easy implementation with other Countries Legislation;
    \item Reference to real PV plant and ESS devices, as presented in the Italian Market;
    \item Adoption of an ESS wear cost model;
    \item Detailed energy purchase cost, with also fixed costs component;
    \item Machine Learning grey-box computation aiming at a XAI optimization model;
\end{itemize}

\noindent In Section \thinspace\ref{sec:hems_architecture}, the REC architecture, with its nodes (MGs) and elements, is depicted while the problem formulation is presented in Section \thinspace\ref{sec:problem_formulation}; details about MG costs (i.e. the ESS wear) are reported in Section \thinspace\ref{sec:mg_costs}; power forecasting ang energy flows optimization tasks are described in Section \thinspace\ref{sec:power_forecasting} and Section \thinspace\ref{sec:optimization_module}, respectively while tests setup and results in Section \thinspace\ref{sec:test_setup} and Section \thinspace\ref{sec:results}, respectively. Finally, conclusions are drawn. 

\vspace{1 cm}

\section{REC Architecture}
\begin{figure}[ht]
  \centering
  \includegraphics[width=0.6\columnwidth]{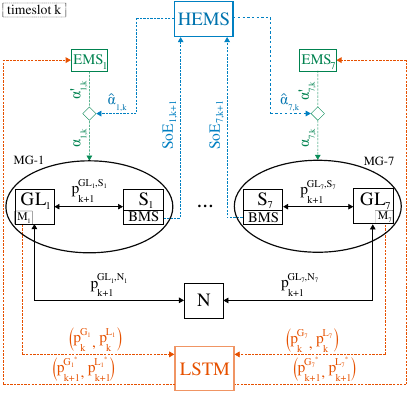}
  \caption{The HEMS architecture.}
  \label{fig:rec_architecture}
\end{figure}

\noindent The REC architecture is pictured in Fig.\thinspace\ref{fig:rec_architecture} in a generic timeslot $k$ of $15$ minutes.

\label{sec:hems_architecture}

\noindent REC nodes, i.e. (MGs), and power flows between them are in black lines; local (node-level) $EMS$ modules and information flows are in green lines while for the $HEMS$ (REC-level) are in blue lines; power prediction $LSTM$ module and information flows are in orange. The REC involves $7$ nodes, enclosed in oval contours, that can exchange power with the external element $N$, the Main Grid. Therefore, each node can work both in isolated and grid-connected mode. Two elements compose each node: the ESS $S_x$ and a composite element named $GL_x$, with $x$ generic node. The latter is the composition of the PV generator (i.e. the PV roof) $G_x$ and the electrical load $L_x$, which comes from the aggregation of the overall household appliances. In turn, both $S_x$ and $G_xL_x$ have a sub-element, the $BMS_x$ or Battery Management System and the meter $M_x$, respectively. The $BMS_x$, which in this work is only simulated, evaluates the $S_x$ State of Energy (SoE) while $M_x$ provides power measures for the timeslot $k$ to the $LSTM$ prediction module.\\ 
\noindent Nearly before the end of the current timeslot $k$, generated and absorbed power values, $P_k^{G_x}$ and $P_k^{L_x}$, are measured by $M_x$ and provided to the $LSTM$ module. In turn, the $LSTM$ predicts the same figures for the next timeslot $k+1$, $P_{k+1}^{G_x^*}$ and $P_{k+1}^{L_x^*}$, and passes them to the $EMS_x$ module. Completely unaware of the other nodes in the REC, the $EMS_x$ performs a power flows optimization having the auto-consumption of $x$ as an objective. In other words, each node plans to act as it was isolated from the other nodes. Optimal decisions $\alpha'_{x,k}$ lead to optimal power flows $P_{k+1}^{GL_x,S_x}$ $P_{k+1}^{GL_x,N_x}$, i.e. the power that flows between $GL_x$ and $S_x$, positive towards $S_x$ and the power that flows between $GL_x$ and $N_x$, positive towards $N_x$. At this step, $\alpha'_{x,k}$ is equal to $\alpha_{x,k}$, since the $HEMS$ has not overwritten the local optimal decision yet, since it will happen in the next phase. Thus, the $BMS_x$ estimates the $SoE_{x,k+1}$. The $SoE$ figures of the overall nodes are passed by their BMSs to the $HEMS$ meaning that the latter acquires information about the state the REC would be in, in $k+1$, if only local auto-consumption would be taken into account. That said, the $HEMS$ performs power flows optimization for the whole REC by aiming at the minimum monetary cost for the REC as an objective. The optimal decision $\hat{\alpha}_{x,k}$ overwrites the local optimal decision $\alpha'_{x,k}$ only if their values are different, since the $HEMS$ is hierarchically superior. Therefore, if $\hat{\alpha}_{x,k}$ is different from $\alpha'_{x,k}$, $\hat{\alpha}_{x,k}$ becomes equal to $\alpha_{x,k}$, the definitive optimal decision for $k+1$; otherwise,  $\alpha_{x,k}$ is equal to $\alpha'_{x,k}$. In the former case, $P_{k+1}^{GL_x,S_x}$ and $P_{k+1}^{GL_x,N_x}$ are recalculated becoming the definitive optimal power flows for $k+1$.

\subsection{Design Choices}
\label{sub-sec:design_choices}
\noindent In this work, many choices are made by adhering to the current European technical legislation ~\cite{red_II}~\cite{iem} as it is inherited by Italy (where the authors come from) ~\cite{legge_8_2020}~\cite{arera_318}~\cite{arera_727}~\cite{regole_GSE}~\cite{mise_2020}~\cite{PNRR}~\cite{decreto_199}~\cite{decreto_200}~\cite{pniec}, henceforth named ``legislation". A detailed explanation of the reasons behind that choices are shown in the following.\\
First of all, a hierarchical EMS for the overall REC energy management whose objective is different from the local auto-consumption is compliant with the legislation that, at the best of our knowledge, aims at cost reduction. Indeed, high PV plant and ESS installation costs could make a green-energy-powered home perspective not affordable for the single customer, from which the idea of customer aggregation comes. In addition, thanks to such an sped up spread of distributed generation, more consumers could be led to rely less and less on the Main Grid, resulting in benefits for the latter. In fact, power peaks inside the electrical infrastructure would be reduced, from which less operational cost to face. In light of that, the Government provides incentives to the RECs participants proportionally to the locally-generated energy they share with each other. Moreover, energy sell is allowed at a fixed price. That way, customers are encouraged to join RECs because they could save on the electricity bill and on installation costs amortization. For what concerns the number of REC nodes, it is derived from data about prototype RECs in Italy, collected by Legambiente ~\cite{legambiente_2022}, an Italian environmental association. More specifically, it turns out that small RECs (equipped with PV generators with overall power between $10$ kWp and $20$ kWp, in line with the target of this work) count between $2$ and $12$ residential participants. Thus the average value of $7$ nodes is adopted. A $15$ minutes long timeslot $k$ is set according to literature ~\cite{kurukuru}. The choice of unifying generator and load accomplishes the legislation ~\cite{red_II}~\cite{legge_8_2020} (Art. $42$-bis, par. $4$), according to which renewable energy consumption should be instantaneous. It means that renewable energy provided by $G_x$ must be primarily consumed by $L_x$ for satisfying the demand, rather than the not completely renewable energy purchased from the Main Grid $N$ ~\cite{link_terna}. To be precise, the legislation states that also renewable energy stored in the ESSs should be consumed instantaneously but it is trivial that $L_x$ should primarily rely on $G_x$ than $S_x$ for avoiding ESSs wear costs. With that premise, $G_x$ and $L_x$ can be represented as a single node $GL_x$. Moreover, there are not power flows between $N_x$ and $S_x$ for the following reasons:

\begin{itemize}
  \item If energy would be purchased from the Main Grid to charge the ESS, the REC would implicitly consider fossil fuel energy as a primary resource while, on the contrary, it must be an emergency resource for energy shortages (renewable energy is primary). Therefore, $S_x$ can be charged only in case of energy overabundance from $GL_x$ and power flows between $N_x$ and $S_x$ towards the latter are not possible; 
  \item Selling energy excess from $GL_x$ is more convenient than selling the same quantity from $S_x$ since, in that case,  ESSs wear costs would occur. Thus, power flows between $N_x$ and $S_x$ towards the former are not possible.
\end{itemize}

\noindent In addition, power flows between $GL_x$ and $N_x$ are bidirectional. In fact, in line with legislation ~\cite{regole_GSE}(par. 6)~\cite{link_prezzo_GSE}, $GL_x$ excess energy can be sold to $N_x$ (other than purchased) at a given price that will be made explicit in Section\thinspace\ref{sec:problem_formulation}. \\

\vspace{ 0.5 cm}

\section{Problem formulation}
\label{sec:problem_formulation}
\begin{figure*}[ht]
  \centering
  \includegraphics[width=\textwidth]{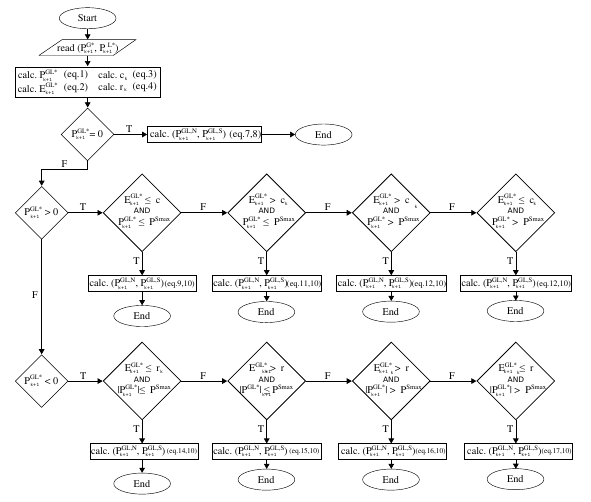}
  \caption{The node level EMS flowchart.}
  \label{fig:node_ems_flowchart}
\end{figure*}

\noindent Node-level $EMSs$ pursue MGs auto-consumption while the $HEMS$ has the REC costs minimization as an objective. Therefore, different problem formulations have to be considered: the node-level problem, tackled by $EMSs$, and the REC-level problem, faced by the $HEMS$. Nevertheless, the formulation below (eq.\thinspace\ref{eq:p_gl}- eq.\thinspace\ref{eq:soe}), whose flowchart is shown in Fig.\thinspace\ref{fig:node_ems_flowchart}, is in common because of the decision variable $\alpha_k$. In fact, as pictured in Fig. \ref{fig:rec_architecture}, the $EMS_x$ sets $\alpha^{'}_{x,k}$ for achieving auto-consumption and the formulation at hand is designed to reach that goal by setting $\alpha_k$ equal to $\alpha^{'}_{x,k}$ that always is equal to $1$. It is worth mentioning that SoE falls into the given bounds implicitly, i.e. without an explicit constraint equation.

\label{sec:node_ems_flowchart}

\begin{equation} 
\label{eq:p_gl}
    P_{k+1}^{GL_x^*}=P_{k+1}^{G_x^*}+P_{k+1}^{L_x^*}  
\end{equation}

\begin{equation}
    E_{k+1}^{GL_x^*}=P_{k+1}^{GL_x^*}+\Delta t  \label{eq:e_gl}
\end{equation}

\begin{equation}
     c_k=Q\thinspace\cdot\thinspace (SoE^{max}-SoE_{x,k})  \label{eq:c_k}
\end{equation}

\begin{equation}
      r_k= Q\thinspace\cdot\thinspace (SoE_{x,k}-SoE^{min}) \label{eq:r_k}
\end{equation}

\begin{equation}
      SoE^{max}= 0.95 \label{eq:soe_max}
\end{equation}

\begin{equation}
      SoE^{min}= 0.15 \label{eq:soe_min}
\end{equation}

\begin{equation}
     P_{k+1}^{GL_x,S_x}=0 \thinspace[kW]\label{eq:p_gl_s}
\end{equation}

\begin{equation}
     P_{k+1}^{GL_x,N}=0 \thinspace[kW] \label{eq:p_gl_n}
\end{equation}

\begin{equation}
     P_{k+1}^{GL_x,S_x}=\alpha_{x,k}\thinspace\cdot\thinspace P_{k+1}^{GL_x^*}  \label{eq:p_gl_s_1}
\end{equation}

\begin{equation}
     P_{k+1}^{GL_x,N}=P_{k+1}^{GL_x^*} - P_{k+1}^{GL_x,S_x}  \label{eq:p_gl_n_1}
\end{equation}

\begin{equation}
     P_{k+1}^{GL_x,S_x}=\alpha_{x,k}\thinspace\cdot\thinspace \frac{c_k}{\Delta t}  \label{eq:p_gl_s_2}
\end{equation}

\begin{equation}
     P_{k+1}^{GL_x,S_x}=\alpha_{x,k}\thinspace\cdot\thinspace P^{S,max}  \label{eq:p_gl_s_3}
\end{equation}

\begin{equation}
     P^{S,max}=7\thinspace [kW]  \label{eq:p_s_max}
\end{equation}

\begin{equation}
    P_{k+1}^{GL_x,S_x}=-\alpha_{x,k} \thinspace\cdot\thinspace |P_{k+1}^{GL_x^*}|  \label{eq:p_gl_s_4}
\end{equation}

\begin{equation}
     P_{k+1}^{GL_x,S_x}=-\alpha_{x,k}\thinspace\cdot\thinspace \frac{r_k}{\Delta\thinspace\dot\thinspace t} \label{eq:p_gl_s_5}
\end{equation}

\begin{equation}
     P_{k+1}^{GL_x,S_x}=-\alpha_{x,k}\thinspace\cdot\thinspace P^{S,max} \label{eq:p_gl_s_6}
\end{equation}

\begin{equation}
\label{eq:soe}
 SoE_{x,k+1}=
\begin{cases}
    \left(
    SoE_{x,k} + \frac{P_{k+1}^{GL_x,S_x} \thinspace\dot\thinspace \Delta t}{Q} \right)\thinspace\cdot\thinspace\eta & \text{if } P_{k+1}^{GL_x^*} > 0 \\
    \left(
    SoE_{x,k} - |\frac{P_{k+1}^{GL_x,S_x} \thinspace\dot\thinspace \Delta t}{Q}| \right) \thinspace\cdot\thinspace\frac{1}{\eta}& \text{otherwise }     
\end{cases}
\end{equation}

\noindent The quantity $ E_{k+1}^{GL_x^*}$  is the amount of energy exchanged in $k+1$ by $GLx$ in a $\Delta t$ of $15$ minutes (one timeslot); $c_k$ is the remaining fraction of the ESS Capacity $Q$ (expressed in terms of energy amount) in $k$, i.e. how much energy the ESS can store yet; $r_k$ is the remaining energy amount in the ESS, in $k$; $SoE^{min}$ and $SoE^{max}$ are the lower and upper bounds of the SoE, respectively, for preserving the ESS health; $P^{(S,max)}$ is the maximum power that the chosen ESS can exchange due to its technical limits (see Sub-section\thinspace\ref{sub-sec:pv_ess_costs} for further details).
\noindent Once the $SoS_{k+1}$ is calculated by the $EMS_x$, it is passed as an input to the $HEMS$ that achieves the optimal values of the decision variable $\hat{\alpha}_{x,k}$
against the REC Objective Function (OF) described by the equations below:

\begin{equation}
      \min_{\alpha_{i,k}} \thinspace(R_{k+1}-C_{k+1}) \label{eq:min}
\end{equation}

\vspace{0.5cm}
\noindent subject to:

\begin{equation}
       \alpha_{i,k}\thinspace\in\thinspace[0,1]\thinspace\thinspace \thinspace \thinspace \thinspace i=1,2,...,n \label{eq:alpha_range}
\end{equation}

\begin{equation}
       n=7 \label{eq:n}
\end{equation}

\begin{equation}
      R_{k+1}=I_{k+1}^{sha} + I_{k+1}^{ret} + I_{k+1}^{sel} \label{eq:r}
\end{equation}

\begin{equation}
      C_{k+1}=h_{k+1}^{ESS} + h_{k+1}^{pur} + h_{k+1}^{ins} \label{eq:c}
\end{equation}

\begin{equation}
      I_{k+1}^{sha}=TP_{REC}\thinspace\cdot\thinspace E_{k+1}^{sha}\label{eq:i_sha}
\end{equation}

\begin{equation}
   I_{k+1}^{ret}=CU_{Af,m}\thinspace\cdot\thinspace E_{k+1}^{sha}\label{eq:i_ret}
\end{equation}

\begin{equation}
\label{eq:i_sel}
 I_{k+1}^{sel}=
\begin{cases}
    PR^3 \thinspace\cdot\thinspace \Delta t \thinspace\cdot\thinspace P_{k+1}^{GL_x,N} & \text{if } P_{k+1}^{GL_x,N} > 0 \\
     0 & \text{otherwise }     
\end{cases}
\end{equation}

\begin{equation}
      TP_{REC} = 110 \thinspace \left[\frac{\text{\euro}}{kWh}\right]\label{eq:te_rec}
\end{equation}

\begin{equation}
       CU_{Af,m} = TRAS_e + max(BTAU_m)  \label{eq:cu_afm}
\end{equation}

\begin{equation}
       TRAS_e = 7,61 \thinspace \left[\frac{\text{\euro}}{kWh}\right] \label{eq:tras_e}
\end{equation}

\begin{equation}
       max(BTAU_m) = 0,61 \thinspace \left[\frac{\text{\euro}}{kWh}\right] \label{eq:max_btau_m}
\end{equation}

\begin{equation}
       E_{k+1}^{sha} = min(E_{k+1}^{gen}, E_{k+1}^{dra}) \label{eq:e_sha}
\end{equation}

\begin{equation}
       E_{k+1}^{gen} = P_{k+1}^{G*} \thinspace\cdot\thinspace \Delta t \label{eq:e_gen}
\end{equation}

\begin{equation}
\label{eq:e_dra}
 E_{k+1}^{dra}=
\begin{cases}
     P_{k+1}^{L*} + P_{k+1}^{GL_x,S_x} & \text{if } P_{k+1}^{GL_x,S_x} > 0 \\
     P_{k+1}^{L*} & \text{otherwise }
\end{cases}
\end{equation}

\begin{equation}
\label{eq:h_pur}
 h_{k+1}^{pur}=
\begin{cases}
     (u_{pur} \thinspace\cdot\thinspace P_{k+1}^{GL_x,N} + u_{pur}^*)\thinspace\cdot\thinspace (1+VAT)& \text{if } P_{k+1}^{GL_x,N} < 0 \\
     u_{pur}^* \thinspace\cdot\thinspace (1+VAT)& \text{otherwise}     
\end{cases}
\end{equation}

\begin{equation}
       h_{k+1}^{ins} = u_{PV}\thinspace\cdot\thinspace|P_{k+1}^{G_x}|\thinspace\cdot\thinspace \Delta t \label{eq:h_ins}
\end{equation}

\begin{equation}
       u_{pur} = 0,212 \thinspace \left[\frac{\text{\euro}}{kWh}\right]  \label{eq:u_pur}
\end{equation}

\begin{equation}
       u_{pur}^* = 0,003 \thinspace \left[\text{\euro}\right]  \label{eq:u_pur*}
\end{equation}

\noindent The HEMS minimizes the difference between total revenues $R_{k+1}$ and total costs $C_{k+1}$ (eq.\thinspace\ref{eq:min}) for the $n$ nodes. $R_{k+1}$ (eq.\thinspace\ref{eq:r}) is the sum of revenues from “shared energy” $I_{k+1}^{sha}$, revenues from the “return of tariff components” (except system charges)” $I_{k+1}^{ret}$ and revenues from energy sold to the Main Grid $I_{k+1}^{sel}$, according to Gestore dei Servizi Energetici (GSE S.p.A), 2023 (par. 6.1). The cost $C_{k+1}$ (eq.\thinspace\ref{eq:c}) comes from the ESS wear cost $h_{k+1}^{ESS}$, the energy purchase from the Main Grid $h_{k+1}^{pur}$ and the overall PV plant and ESS installation cost $h_{k+1}^{ins}$. Revenues are calculated thanks to the “premium tariff” for RECs $TP_{REC}$, the “unit consumption of the monthly flat rate” $CU_{Af,m}$, the energy sell cost $PR_{3}$, the “tariff for the broadcast service” $TRAS_{e}$ and the “maximum value of the distribution component variable” $max(BTAU_{m})$, according to Gestore dei Servizi Energetici (GSE S.p.A), 2023 and Autorità di Regolazione per Energia Reti e Ambiente (ARERA), 2020-2022. The “shared energy” $E_{k+1}^{sha}$ is a very important quantity since it is the main revenue factor for the REC. According to Italian Government, 2020b (Law 28th February 2020 n°8):

\noindent 
\begin{center}
\textit{The shared energy is equal to the minimum, in each hourly period, between the electricity produced and fed into the grid by the plants to renewable sources and the electricity taken from the whole associated end customers.}
\end{center}

\noindent Consistently, eq.\thinspace\ref{eq:e_sha}  formulates that principle with $E_{k+1}^{gen}$ and $E_{k+1}^{dra}$ the amount of energy generated by PV plant and the amount of energy the REC absorbs or stores in the ESS, in $k+1$. Costs are calculated through $u_{pur}$ energy purchase unit price from the Main Grid, $u_{pur}^*$ energy purchase fixed price from the Main Grid, $u_{PV}$ PV plant installation cost coefficient, $h_{k+1}^{ESS}$ ESS wear cost, being $d$ the number of timeslots. Those prices and costs are derived and discussed in Section\thinspace\ref{sec:mg_costs}. 

\vspace{1 cm}

\section{MG costs}
\label{sec:mg_costs}

\noindent The choices about costs and prices are described in this Section. Together with the energy purchase price from the Main Grid, PV plant installation cost as well as the ESS operating wear cost model are laid out and discussed.

\subsection{Energy purchase price}
\label{sub-sec:energy_price}
\noindent Energy price in Italy is composed by a unit price (per $kWh$) and a fixed price. Both of them are derived from the economic conditions of an Italian energy operator, fund using the ARERA portal ~\cite{ARERA_portal}. The unit price is $0.212 \thinspace \frac{\text{\euro}}{kWh}$ while the fixed price, for a single node in the generic timeslot, is $0.003 \thinspace \text{\euro}$. Those economic conditions refer to $6 kW$ clients, derived from the average max power absorbed by nodes in the dataset time horizon (year 2019), (shown in Sub-Section \thinspace \ref{subsec:dataset_presentation}). Those costs are increased by the Italian VAT ($10\thinspace\%$).

\subsection{PV and ESS wear costs}
\label{sub-sec:pv_ess_costs}

\noindent The PV plant wear cost $h_{ins}$, with reference te eq.\thinspace\ref{eq:h_ins}, is estimated proportionally to the usage, i.e. to the amount of energy exchanged between the generator and the rest of the node elements. The coefficient $u_{PV}$, expressed in $\text{\euro}/kWh$, is the average ratio, among the $7$ nodes, between the PV installation cost and the overall energy generated during the whole year 2019 (the dataset at our disposal. The PV installation cost is about $6.000\thinspace \text{\euro}$  ~\cite{ENEL_PV_3} for a $3 \thinspace kW_p$ PV plant while about $7.400\thinspace \text{\euro}$ ~\cite{ENEL_PV_4} for a $4 \thinspace kW_p$ PV plant, according to the store Enel X, an important Italian Company. Those costs include also inspection, design and VAT. 

\noindent The ESS wear cost comes from a proper wear cost model based on ~\cite{han_2018}. The model formulation is presented below:

\begin{equation}
      h_{k+1}^{ESS} = \frac{\Delta t}{2} \thinspace \cdot \thinspace (W_{SoE_{x,k}} + W_{SoE_{x,k+1}})\thinspace \cdot \thinspace|P_{k+1}^{GL_x,S_x}| \label{eq:ess_cost}
\end{equation}

\vspace{0.5cm}
\noindent with:

\begin{equation}
       W_{SoE_{x,k}} = \frac{u_{ESS}}{2\thinspace \cdot \thinspace Q \thinspace \cdot \thinspace \eta} \thinspace \cdot \thinspace \frac{b \thinspace \cdot \thinspace (1-SoE_{x,k})^{(b-1)}}{a} \label{eq:density_w_soe}
\end{equation}

\noindent The ESS installation price $u_{ESS}$ is about $5.000\thinspace \text{\euro}$ and, as for the PV plant, it includes also inspection, design and VAT costs. It is a $5\thinspace kWh$ battery ~\cite{ENEL_ESS} whose efficiency $\eta$ is $0.98$ and whose maximum output power $P^{S,max}$ is $7 \thinspace kW$ (Tab.\thinspace \ref{tab:ess}). 

\renewcommand{\arraystretch}{1.5}%
\begin{table}[ht]
  \centering
  \caption{ESS main features.}
  \begin{tabular}{|c|c|}
  \hline
  \textbf{Feature} & \textbf{Value} \\
  \hline
   Price & $5000\thinspace \text{\euro}$\\
  \hline
   Capacity & $5\thinspace kWh$\\
  \hline
   Peak power & $7\thinspace kW$\\
  \hline
   Efficiency ($\eta$) & $0.98$\\
  \hline
  \end{tabular}
  \label{tab:ess}
\end{table}

\begin{figure}[ht]
  \centering
  \includegraphics[width=0.5\columnwidth]{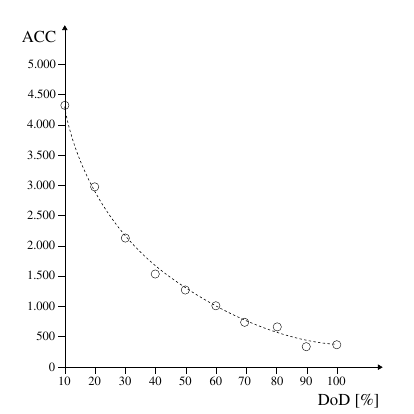}
  \caption{An example of ACC-DoD Lithium-Ions battery experimental curve.}
  \label{fig:acc}
\end{figure}

\noindent The quantity $h_{k+1}^{ESS}$, i.e. the ESS wear cost in $\text{\euro}$, is calculated through $W_{SoE}$, in $\frac{\text{\euro}}{kWh}$, that is defined in ~\cite{han_2018} as the ``Average Wear Cost” and can be interpreted as a cost density, since it represents the cost at given $SoE$ value. Moreover, $a$ and $b$ are Lithium-Ions battery parameters related to the specific ESS and precisely to the Achievable Cycle Count (ACC) and the Depth of Disharge (DoD) experimental curve interpolation (Fig.\thinspace\ref{fig:acc}, eq.\thinspace\ref{eq:acc}). 

\begin{equation}
      ACC(DoD_{x,k}) =\frac{a}{DoD_{x,k}^b}  \label{eq:acc}
\end{equation}

\noindent In other words, $a$ and $b$ values control the shape of the aforementioned curve, i.e. they provide information about the ESS lifespan. Therefore, $h_{k+1}^{ESS}$ is the cost of the ESS for changing its $SoE$ lavel from $SoE_{x,k}$ to $SoE_{x,k+1}$, due to a power exchange in any direction $|P_{k+1}^{GL_x,S_x}|$.

\noindent Since ENEL X does not provide an ACC-DoD experimental curve for the chosen ESS, $a$ and $b$ are rationally derived from literature ~\cite{han_2018}~\cite{kim_2019}, whose figures are reported in Tab. \ref{tab:ACC_a_b}. To be in line with literature, $a$ and $b$ values are $694$ and $0.795$, respectively.

\renewcommand{\arraystretch}{1.5}%
\begin{table}[ht]
  \centering
  \caption{ESS $a$ and $b$ values for different batteries, according to literature.}
  \begin{tabular}{|c|c|c|}
  \hline
  \textbf{Source} & \textbf{a} & \textbf{b} \\
  \hline
   ~\cite{han_2018} & 694 & 0.795\\
  \hline
  ~\cite{kim_2019} & 1131 & 1.825\\
  \hline
  ~\cite{kim_2019} & 2744 & 1.665\\
  \hline
  \end{tabular}
  \label{tab:ACC_a_b}
\end{table}

\vspace{1cm}

\section{Power Forecasting}
\label{sec:power_forecasting}
\noindent Thanks to power generation and power consumption historical data (described more in deep in Subsection \ref{subsec:dataset_presentation}), a Long Short-Time Memory (LSTM) Neural Network is trained for power forecasting.  At each timeslot, the module predicts power figures for the next timeslot, in an Open Loop fashion. That Open Loop choice is made for improving prediction performance, under the assumption that a proper meter measures and provides the power value in $k$ to the LSTM. That is in line with the EMSs and HEMS optimal decisions that are taken timeslot after timeslot, in an online the energy management approach.

\subsection{Dataset presentation}
\label{subsec:dataset_presentation}
\begin{figure}[ht]
  \centering
  \includegraphics[width=0.7\columnwidth]{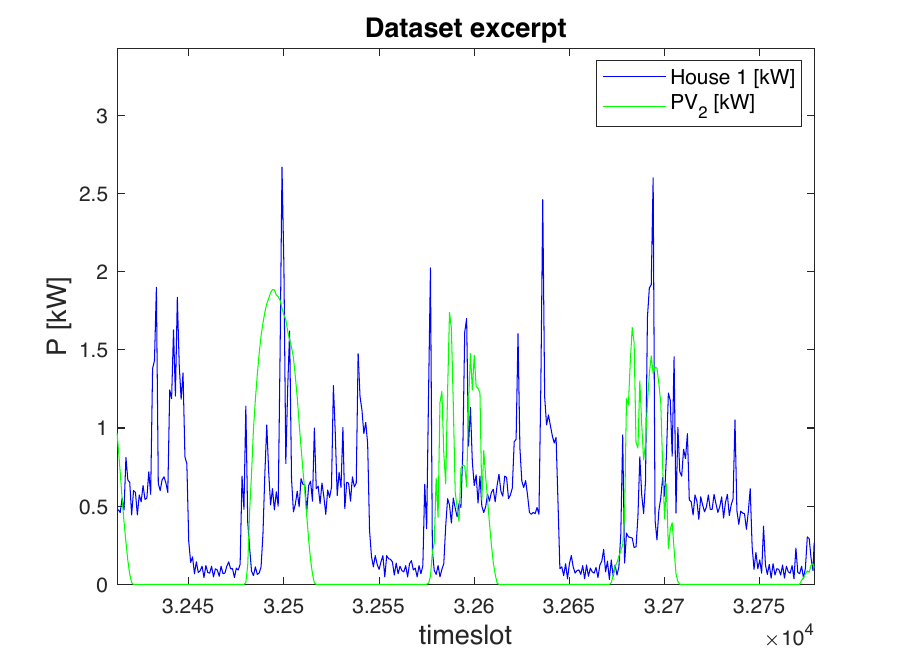}
  \caption{Dataset excerpt with one house and one PV plant power profiles.}
  \label{fig:dataset_excerpt}
\end{figure}

\noindent The dataset comes from an online open data supply belonging to the OpenAIRE European Project. It counts power generation figures for $3$ residential PV plants (one $3$ kWp, named \textit{PV 2} and two $4$ kWp each, named \textit{PV 1} and \textit{PV 3}) together with power consumption measures for $15$ residential homes (each of them named \textit{Home}), for the whole year 2019. Data resolution is $15$ minutes, as required for this work. Power consumption of $7$ houses, chosen randomly among the overall $15$ ones, are included into the computation while the $3$ generation profiles are randomly associated to them. An example of power generation and consumption profiles is provided in Fig.\thinspace \ref{fig:dataset_excerpt} and some additional information are reported in Tab. \ref{tab:dataset_info}.

\renewcommand{\arraystretch}{1.5}%
\begin{table}[ht]
  \centering
  \caption{Dataset information.}
  \begin{tabular}{|c|c|c|c|}
  \hline
  \textbf{Profile} & \textbf{Min [kW]} & \textbf{Max [kW]} & \textbf{Mean [kW]}\\
  \hline
    PV $1$ & $0.000$ & $3.472$ & $0.600$\\
  \hline  
    PV $2$ & $0.000$ & $2.797$ & $0.488$\\
  \hline
    PV $3$ & $0.000$ & $3.439$ & $0.433$\\
  \hline
    Home $1$ & $0.020$ & $3.784$ & $0.362$\\
  \hline
    Home $2$ & $0.008$ & $4.980$ & $0.256$\\
  \hline
    Home $3$ & $0.016$ & $6.324$ & $0.362$\\
  \hline
    Home $4$ & $0.004$ & $7.044$ & $0.663$\\
  \hline
    Home $5$ & $0.028$ & $5.852$ & $0.426$\\
  \hline
    Home $6$ & $0.024$ & $6.324$ & $0.556$\\
  \hline
    Home $7$ & $0.016$ & $5.864$ & $0.441$\\
  \hline
  \end{tabular}
  \label{tab:dataset_info}
\end{table}

\subsection{Model details and workflow}
\label{subsec:workflow_and_model_parameters}
\noindent The forecasting model, (an LSTM implemented through MathWorks\textsuperscript{\textregistered} Matlab libraries), is trained using a $328$ days training set , i.e. the $90\%$ of the days in the whole year 2019 (the overall dataset). Ergo, the test set counts $37$ days of power figures ($10\%$ of the dataset). The LSTM training options and meta-parameters are laid-out in Tab \thinspace \ref{tab:lstm}.

\renewcommand{\arraystretch}{1.5}%
\begin{table}[ht]
  \centering
  \caption{LSTM training options and meta-parameters.}
  \begin{tabular}{|c|c|}
  \hline
  \textbf{Figure} & \textbf{Value}\\
  \hline
    LSTM layers& 128 \\
  \hline
    Max epochs& 200 \\
  \hline
    Shuffle& $every-epoch$ \\
  \hline
    Solver & $RMS Propagation$ \\
  \hline
    Gradient Decay & $0.90$ \\
  \hline
    Squared Gradient Decay & $0.99$ \\
  \hline
    Epsilon & $1.00e^{-08}$ \\
  \hline
    Initial Learning Rate & $1.00e^{-03}$ \\
  \hline
   Learning Rate Schedule & $none$ \\
  \hline
   Learning Rate Drop Factor & $0.10$ \\
  \hline
   Learning Rate Drop Period & $10$ \\
  \hline
   L2 Regularization & $1.00e^{-04}$ \\
  \hline
   Gradient Threshold Method & $l2norm$ \\
  \hline  
   Gradient Threshold & $Inf$ \\
  \hline  
  \end{tabular}
  \label{tab:lstm}
\end{table}

\noindent Roughly at the end of the current day, the LSTM updates its $state$ using power predictions for the previous $35$ days. Consequently, the $state$ update occurs every $24$ hours while training is made only once on historical data. As aforementioned, at each $k$ the LSTM predicts power figures for $k+1$ in an Open Loop fashion. Succintly, an Open Loop prediction for $k+1$ is performed by considering the $state$ at $k$ like in Closed Loop fashion but, differently, also by taking the real power figure in $k$ as an input rather than its prediction. Better precision comes from Open Loop predictions, since real figures are involved in the workflow but a proper meter is assumed to be installed for measuring that figures and providing them to the LSTM at each timeslot.
\noindent Average computational costs for training, state update and prediction itself, between the overall predictions (homes and PV plants) are displayed in Tab. \thinspace \ref{tab:computational_costs_LSTM}. The model runs on $12^{th}$ Gen Intel(R) Core(TM) i9-12900K workstation with $3.19$ GHz processor and $128$ GB RAM. 

\renewcommand{\arraystretch}{1.5}%
\begin{table}[ht]
  \centering
  \caption{LSTM computational costs.}
  \begin{tabular}{|c|c|}
  \hline
  \textbf{Figure} & \textbf{Value [s]}\\
  \hline
    Training & $180$ \\
  \hline
    $state$ update& $0.310$ \\
  \hline
    Test (prediction)& $0.003$ \\
  \hline
  \end{tabular}
  \label{tab:computational_costs_LSTM}
\end{table}

\subsection{Forecasting results}
\label{subsec:forecasting_results}
\noindent Power predictions are represented in Tab.~\ref{tab:power_predictions} for PV plant $1$ and home $1$ by way of example. Predictions are made for $96$ timeslots or $24$ h for a wider overview on the figures, keeping in mind a timselot-by-timeslot forecasting is required.

\begin{table*}[ht]
  \centering
  \begin{tabularx}{\textwidth}{|X|X|}
    \hline
    \includegraphics[width=\linewidth]{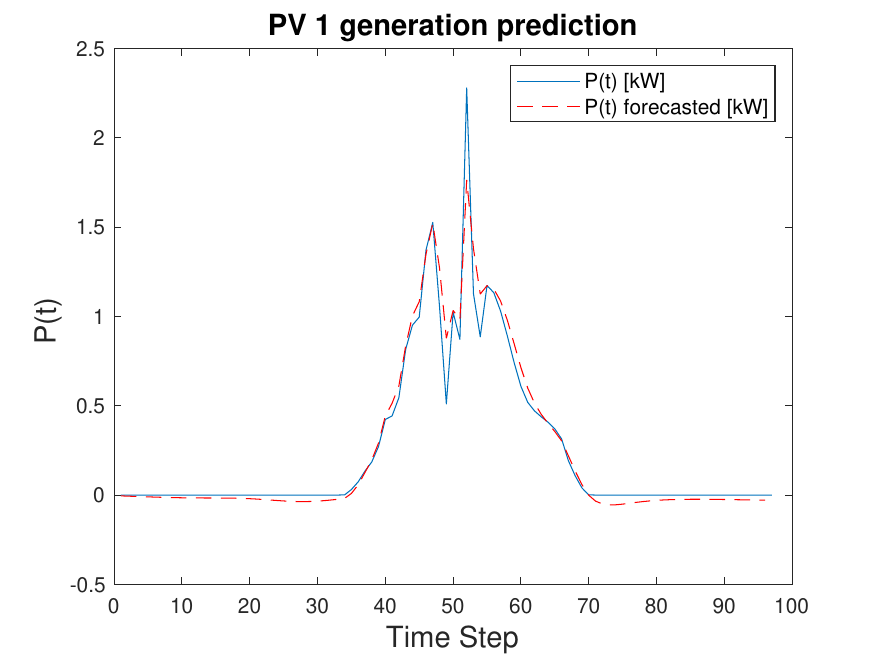} &
    \includegraphics[width=\linewidth]{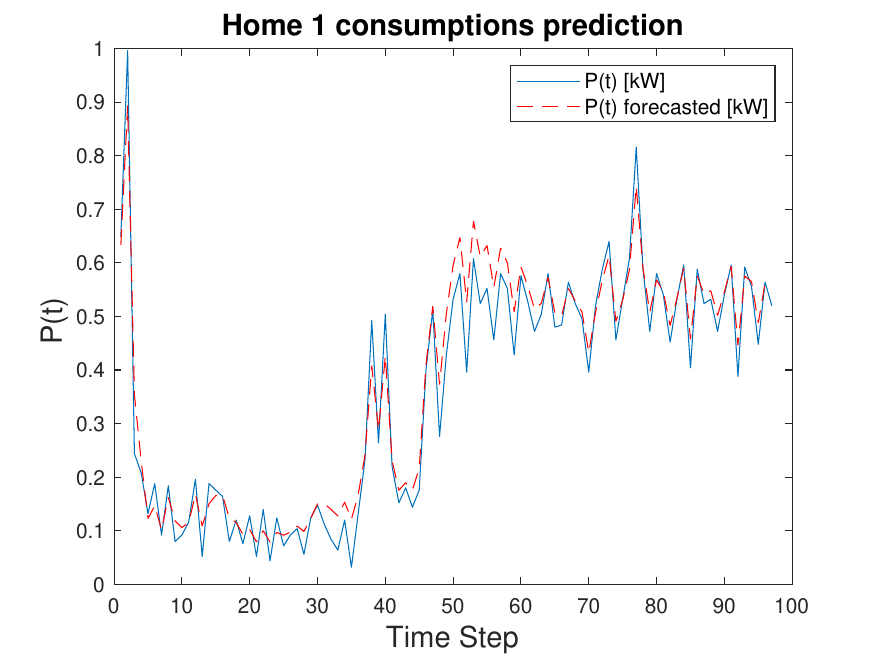} \\
    \hline
  \end{tabularx}
  \captionsetup{skip=10pt}
  \caption{Power predictions.}
  \label{tab:power_predictions}
\end{table*}

\noindent RMSE is calculated as the average between the single-timeslot predictios RMSE values. Figures are given in Tab.\thinspace \ref{tab:rmse}.

\renewcommand{\arraystretch}{1.5}%
\begin{table}[ht]
  \centering
  \caption{Power predictions RMSE values.}
  \begin{tabular}{|c|c|}
  \hline
  \textbf{Prediction} & \textbf{RMSE [kW]}\\
  \hline 
  PV plant $1$ & $0.089$\\
  \hline 
  PV plant $2$ & $0.057$\\
  \hline
  PV plant $3$ & $0.058$\\
  \hline
  Home $1$ & $0.088$\\
  \hline
  Home $2$ & $0.058$\\
  \hline
  Home $3$ & $0.119$\\
  \hline
  Home $4$ & $0.192$\\
  \hline
  Home $5$ & $0.162$\\
  \hline
  Home $6$ & $0.166$\\
  \hline
  Home $7$ & $0.130$\\
  \hline
  \end{tabular}
  \label{tab:rmse}
\end{table}

\noindent Figures are in line with literature, as discussed below. 
\noindent In ~\cite{forecast_1}, LSTM PV power prediction are performed for $21$ PV plants that range from $100 \thinspace kW_p$ to $8500 \thinspace kW_p$. RMSE values fall between $0.04 \thinspace kW$ and $0.112 \thinspace kW$ with an average of $0.07 \thinspace kW$. Such a result could suggest that the RMSE is roughly independent of the PV plant nominal power. Quite similar figures are laid out in ~\cite{forecast_2}, where an LSTM leads to an RMSE of $0.16 \thinspace kW$ for a MG PV plant of $4\thinspace  kW_p$ (a plant size adopted in this work). That said, since the average RMSE for the PV prediction in this work is $0.068 \thinspace kW$  (Tab. \ref{tab:rmse}, it can be stated that those are good results. 
\noindent As concerns load forecasting, in ~\cite{forecast_3}, a minimum RMSE of $0.79 \thinspace kW$ results for an house load profile (which spans from roughly $0 \thinspace kW$ up to $8 \thinspace kW$), among different NN-based prediction approaches. In ~\cite{forecast_4}, an LSTM model leads to an RMSE of about $3\thinspace kW$ for a power profile that fluctuates roughly between $17 \thinspace kW$ and $80 \thinspace kW$. In ~\cite{forecast_5}, an LSTM predictor makes an RMSE error of about $200 \thinspace MW$ for and aggregated load profile that spans from approximately $1000 \thinspace MW$ up to $5000 \thinspace MW$. Moreover, in ~\cite{forecast_6}, the RMSE for an LSTM load forecast for a single household appliance (electric heater) is about $9 \thinspace W$, with absorbed power that falls between $0 \thinspace W$ and $90 \thinspace W$. Since load profiles means are not provided for the above-mentioned works, peak absorbed power is taken as a reference value for results comparisons, representing a measure of the load profile size. Thus, is turns out that a RMSE-peak power ratio between $0.04$ and $0.1$ is acceptable, according to literature. For the load predictions in this work, that rations falls between $0.01$ and $0.03$ with an average of $0.02$. That said, predictions are reasonably good.

\vspace{1 cm}

\section{Optimization Module}
\label{sec:optimization_module}
\noindent The optimization module refers only to the $HEMS$, since the single MG EMSs belong to auto-consumption, according to the problem forumlation described in \ref{sec:problem_formulation}. The module relies on a Fuzzy Inference System - Genetic Algorithm (FIS-GA) paradigm where the GA encodes the FIS parameters (i.e. Antecedents, Consequents, Rule Weights and Membership Function - MFs - abscissas) and optimizes them against the overall REC OF (eq. \thinspace \ref{eq:min}).

\subsection{Optimization workflow}
\label{subsec:optimization_workflow}
\noindent The GA encodes the FIS parameters so that a single Individual of the Populations is generated for each FIS model. In other words, the Population counts a set of solutions to the problem at hand each of which is a FIS that takes the MGs $SoEs$ as an input and returns the OF values. Thanks to its operators, the GA makes Generations pass and the FIS that encodes the best solution comes as an output.

\subsection{FIS structure}
\label{subsec:fis-encoding}
\begin{figure}[ht]
  \centering
  \includegraphics[width=0.45\columnwidth]{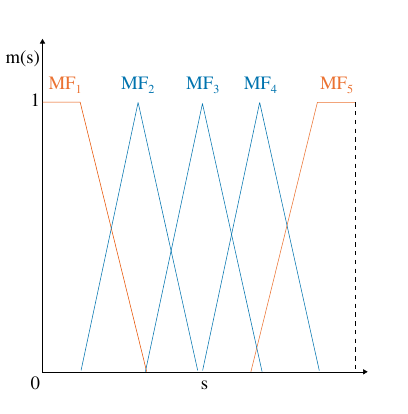}
  \caption{The FIS Term Set.}
  \label{fig:term_set}
\end{figure}

\noindent The FIS is a one-input one-output Mamdamy-type model with a Term Set composed by $5$ MFs, both trapezoidal and triangular, as represented in Fig.\thinspace\ref{fig:term_set}. Thus, the Rule Set counts $5$ Rules and the overall MFs are $25$.

\noindent Therefore, the overall $SoE$ values for $k+1$ are passed as input to the FIS and fuzzyfied by the same input Term Set. Likewise, $\hat{\alpha}_{.,k}$ outputs are de-fuzzified by a single output Term Set.
\noindent With the aim of reducing the variables number, MFs abscissas come from the formulation presented in [capillo] that makes it possible to have only $2$ MF parameters for shaping both a triangular and a trapezoidal MF. For the sake of completeness, that formulation is reported below together with a proper graphic contribution (Fig.\thinspace\ref{fig:mf_encoding}):

\begin{figure}[ht]
  \centering
\includegraphics[width=0.35\columnwidth]  {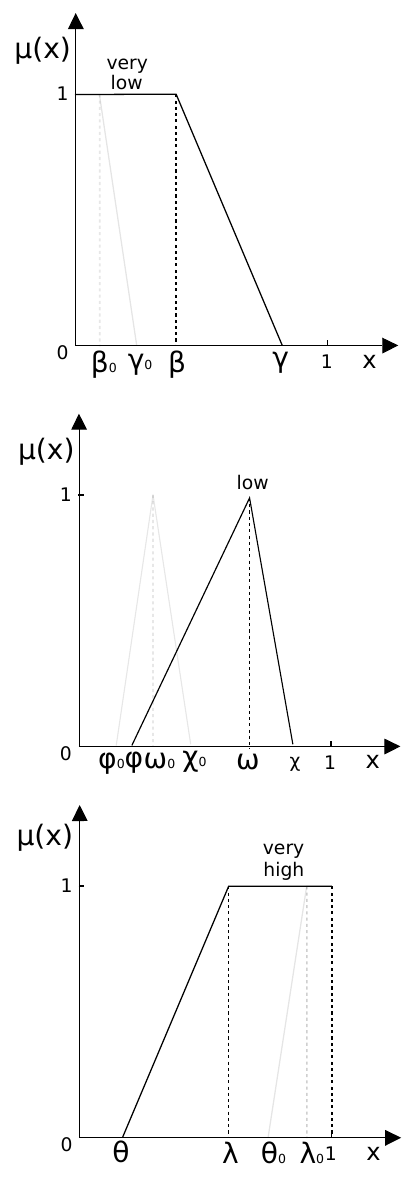}
  \caption{MFs encoding.}
  \label{fig:mf_encoding}
\end{figure}

\begin{equation}
\label{eq:encoding_1}
\gamma = g^{'}_{very \thinspace low} \thinspace \gamma_0 
\end{equation}
\begin{equation}
\label{eq:encoding_2}
\beta = g^{''}_{very \thinspace low} \thinspace \gamma 
\end{equation}
\begin{equation}
\label{eq:encoding_3}
\theta =g^{'}_{very \thinspace high} \thinspace (1 - \theta_0) 
\end{equation}
\begin{equation}
\label{eq:encoding_4}
\lambda = \theta + g^{''}_{very \thinspace high} \thinspace (1-\theta) 
\end{equation}

\begin{equation}
\begin{aligned}
\label{eq:encoding_5}
\phi =
  \begin{cases}
  \phi_0 - (\frac{g^{'}_{low}}{2} \thinspace L_{low} - \frac{L_tr}{2})    &  if \hspace{0.3 cm} g^{'}_{low} \geq 1
  \\
 \phi_0 + (- \frac{g^{'}_{low}}{2} \thinspace L_{low} + \frac{L_tr}{2})  & if \hspace{0.3 cm} 0.01 \geq g^{'}_{low} < 1
\end{cases}
\end{aligned}
\end{equation}

\begin{equation}
\begin{aligned}
\label{eq:encoding_6}
\xi =
  \begin{cases}
  \xi_0 + (\frac{g^{''}_{low}}{2} \thinspace L_{low} - \frac{L_tr}{2})    &  if \hspace{0.3 cm} g^{'}_{low} \geq 1
  \\
  \xi_0 - (- \frac{g^{''}_{low}}{2} \thinspace L_{low} + \frac{L_tr}{2})  & if \hspace{0.3 cm} 0.01 \geq g^{'}_{low} < 1
\end{cases}
\end{aligned}
\end{equation}

\begin{equation}
\label{eq:encoding_7}
\omega = \phi + g^{''}_{low} \thinspace \frac{(\xi - \phi)}{2}
\end{equation}

with:

\begin{equation}
\label{eq:encoding_8}
\gamma_0 = 0.25
\end{equation}

\begin{equation}
\label{eq:encoding_9}
\theta_0 = 0.75
\end{equation}

\begin{equation}
\label{eq:encoding_10}
0.04 \geq g^{'}_{very \thinspace low} \leq 4.00
\end{equation}

\begin{equation}
\label{eq:encoding_11}
0.01 \geq g^{''}_{very \thinspace low} \leq 0.99
\end{equation}

\begin{equation}
\label{eq:encoding_12}
0.04 \geq g^{'}_{very \thinspace high} \leq 4.00
\end{equation}

\begin{equation}
\label{eq:encoding_13}
0.01 \geq g^{''}_{very \thinspace high} \leq 0.99
\end{equation}

\begin{equation}
\label{eq:encoding_14}
0.01 \geq g^{'}_{low} \leq \frac{1}{L_{tr}}
\end{equation}

\begin{equation}
\label{eq:encoding_15}
0.01 \geq g^{''}_{low} \leq 1.99
\end{equation}

\noindent where $\gamma$, $\beta$, $\theta$, $\lambda$, $\phi$, $\xi$ and $\omega$ are the MFs abscissas (Fig. \ref{fig:mf_encoding}) and $\gamma_0$, $\beta_0$, $\theta_0$, $\lambda_0$, $\phi_0$, $\xi_0$ and $\omega_0$ their default values. The quantities designated by $g$ are the GA Genes that encode the FIS structure and that will be discussed in the next Sub-section.

\subsection{GA Encoding}
\label{subsec:ga_encoding}

\noindent The generic GA Individual encodes the Mamdami FIS structure as shown below:

\begin{figure}[ht]
  \centering
  \includegraphics[width=0.55\columnwidth]{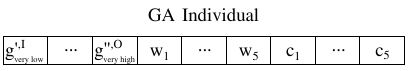}
  \caption{GA Individual encoding.}
  \label{fig:individual}
\end{figure}

\noindent The Individual counts $30$ Genes: $20$ MFs abscissa Genes, $5$ Weights Genes and $5$ Consequents Genes. More precisely, each MF is shaped thanks to two Genes, $g'$ and $g''$ (eq. \thinspace\ref{eq:encoding_10}-\ref{eq:encoding_15}, that leads to overall $10$ Genes for the input Term Set and $10$ Genes for the output Term Set. Weights and Consequents Genes number is equal to the number of Rules.

\vspace{1 cm}

\section{Training and Test Setup}
\label{sec:test_setup}
\noindent Tests are carried on both on the GA as standalone algorithm against benchmark problems and on the FIS-GA algorithm against the REC energy management problem at hand. The setup of the aforementioned tests is discussed in this Section.

\subsection{GA Preliminary Tests}
\label{subsec:GA_tests_setup}
\noindent The GA is preliminary tested on the most common benchmark OFs according to literature. Optimization is performed against two variables OFs \textit{Spherical}, \textit{Rastrigin}, \textit{Rosenbrock}, \textit{Swhefel} and \textit{Griewank}. Solutions and optimal OF values are averaged between $10$ GA executions and Standard Deviation is calculated complying with the Evolutionary Computation stochastic nature.

\subsection{FIS-GA Training and Test}
\label{subsec:FIS-GA_training_test_setup}

\noindent The FIS-GA is trained on a half day timeslots. That choice is in agreement with literature, according to which efforts are made towards a training based on a small set of historical data ~\cite{parking_lots}. Moreover, a FIS-based model like the one at hand implies a Rule Base building process so that training on a half day, which involves differents configurations (no PV production and peak PV production) leads to satisfying results. Training results are averaged between $10$ executions, due to the stochastic nature of the GA optimizer. For the same reason, also the model computational costs are averaged between executions. The best trained model among the aforesaid ten is then tested as explained in the following.

\noindent Conforming to literature ~\cite{parking_lots}, two different test setups are contemplated:

\begin{itemize}
  \item \textbf{Offline test}: the FIS-GA optimizes the REC energy flows for the next $24$ h by taking as an input the overall next-day power predictions at once; 
  \item \textbf{Online tests}:the FIS-GA optimizes the REC energy flows for the next timeslot $k+1$ by taking as an input the next-timeslot power predictions one by one, i.e. timeslot-by-timeslot. 
\end{itemize}

\noindent In the first case, a perfect knowledge about future PV generation is taken for granted. That is a far-fetched statement because the HEMS relies on power forecasts (an estimation of actual values) and also because that forecasts, without an Open-Loop timeslot-based approach, would be less accurate timeslot after timeslot. Nevertheless, that setup is aimed at assaying the optimizer precision against a complex problem even though, as just discussed, the optimization problem to solve is not realistic. A proper benchmark algorithm is chosen for evaluating the offline solution. It consists of a first-step optimization performed by the MathWorks\textsuperscript{\textregistered} Matlab \textit{ga} solver and a second-step optimization performed by the MathWorks\textsuperscript{\textregistered} Matlab \textit{fmincon} solver. Such a procedure is made for facing the problem complexity by narrowing the OF domain (first-step) and then delegating an exact algorithm for finding the benchmark solution (second-step). The best trained model over the $10$ executions is applied on the test set.
\noindent As concerns the \textit{Online tests}, they can be considered as simulation of the real HEMS application. Like in the previous case, there is a similar assumption about the power predictions measurement but it is limited to the timeslots the tests are carried on rather than the overall next-day timeslots. For both of the two setup, the auto-consumption solutions on the test set are also achieved and reported. That is a very important benchmark for the $HEMS$, since the solution of the latter must be better than the auto-consumption one in order to justify its usefulness. Moreover, training is performed $10$ times, due to the stochastic nature of the GA, so that results are averaged and standard deviations are achieved. In order to achieve generic results, i.e. not focused on a specific REC configuration, an initial $SoE$ value of $0.5$ is set for each node.

\noindent The GA meta-parameters and operators adopted for the tests above are reported in Tab. \thinspace \ref{tab:ga_settings}:

\renewcommand{\arraystretch}{1.5}%
\begin{table}[ht]
  \centering
  \caption{GA meta-parameters and operators.}
  \begin{tabular}{|c|c|}
  \hline
  \textbf{Figure} & \textbf{Value}\\
  \hline 
  Population & $100$\\
  \hline
  Cross. Fract. & $0.7$\\
  \hline
  Mut. Prob. & $0.5$\\
  \hline
  Crossover & $convex$\\
  \hline
  Mutation & $uniform$\\
  \hline
  Stop. Cond. & $max gen.$\\
  \hline
  Max Gen. & $50$\\
  \hline
  \end{tabular}
  \label{tab:ga_settings}
\end{table}

\vspace{1 cm}

\section{Results}
\label{sec:results}

\noindent Preliminary GA tests results are reported in Tab.\thinspace \ref{tab:benchmark_ofs}. Being the mean percentage error both for solution an OF ($err_{sol}$ and $err_{OF}$) are under the $5$\% for all the benchmark problems, the GA is reliable, at the best of our knowledge. In addition, given that the standard deviation both on the solution and the optimal OF value is very low, the GA is considered robust, at least for the presented benchmark problems. 

\begin{table*}
  \centering
  \caption{FIS-GA online test results.}
  \begin{tabular}{|c|c|c|p{2.5cm}|c|c|c|}
  \hline
  \centering
  \textbf{Task} & 
  \centering
  \textbf{Auto-cons. OF  [\text{\euro}] }  &
  \centering \textbf{Offline OF [\text{\euro}]} & 
  \centering
  \textbf{Online OF  [\text{\euro}] }  &
  \textbf{$\epsilon'$  [\%]} &
  \textbf{$\epsilon''$  [\%]} &
  \textbf{$z$ [\%]}\\
  \hline  
  \centering
  \textit{Online test} &42.02&\centering32.67&\centering33.71&3.18&5.11&-19.77\\
  \hline 
  \end{tabular}
  \label{tab:hems_online}
\end{table*}

\renewcommand{\arraystretch}{1.5}%
\begin{table}[ht]
  \centering
  \caption{GA solutions against benchmark OFs, averaged between $20$ executions.}
  \begin{tabular}{|c|c|C{1.9 cm}|}
  \hline
  \centering \textbf{Benchmark} & \centering \textbf{OF bench. }&
  \textbf{OF test GA\newline (std. dev.)}\\
  \hline 
  \centering
  \textit{Spherical} &0.000 &2.034E-07 \quad \quad(3.125E-13)\\
  \hline 
  \centering
  \textit{Rastrigin} &0.000&8.063E-09 (8.131E-16)\\
  \hline
  \centering
  \textit{Rosenbrock} & 0.000& 0.000 (8.086E-06)\\
   \hline
  \centering
  \textit{Schwefel} &0.000&2.546E-05 (8.714E-16)\\
  \hline 
  \centering
  \textit{Rastrigin} & 0.000 &8.063E-09 (8.131E-16)\\
  \hline 
  \centering
  \textit{Griewank} & 0.000&0.006 (3.479E-05)\\
  \hline
  \end{tabular}
  \label{tab:benchmark_ofs}
\end{table}

\noindent The FIS-GA training and offline test results are laid out in Tab.\thinspace\ref{tab:hems_train_offline}.

\begin{table}[ht]
  \centering
  \caption{FIS-GA training offline test results.}
  \begin{tabular}{|c|c|p{1.5cm}|c|}
  \hline
  \centering
  \textbf{Task} & 
  \centering
  \textbf{OF b. [\text{\euro}]} &
  \centering \textbf{OF F. [\text{\euro}] ($\sigma$)} & 
  \textbf{$\epsilon$ [\%]}\\
  \hline 
  \centering
  \textit{Train.} &12.23&\centering12.54 \quad \quad\quad(0.18)&2.50\\
  \hline 
  \centering
  \textit{Off. test} &32.07&\centering32.67&1.87\\
  \hline 
  \end{tabular}
  \label{tab:hems_train_offline}
\end{table}

\noindent With $\epsilon$ being the percentage error made by the FIS-GA if compared to the benchmark. In addition, the local auto-consumption optimal OF, returned by the benchmark algorithm, is $42.02 \thinspace \text{\euro}$. The online FIS-GA performance are displayed in Tab.\thinspace\ref{tab:hems_online}, compared to the offline and the auto-consumption OFs. The quantity $\epsilon'$ is the error of the online optimal optimal OF compared to offline optimal OF, while $\epsilon''$ is the error of the online optimal OF compared to the benchmark optimal OF (shown in Tab.\thinspace \ref{tab:hems_train_offline}). Moreover, $z$ represents the percentage savings of the online approach compared to the auto-consumption. The model computational costs are reported in Tab.\thinspace \ref{tab:computational_costs}.

\renewcommand{\arraystretch}{1.5}%
\begin{table}[ht]
  \centering
  \caption{Model computational costs.}
  \begin{tabular}{|c|c|}
  \hline
  \textbf{Cost type} & \textbf{Value}\\
  \hline 
  Training & $46\thinspace min$\\
  \hline 
  Offline test & $0.45\thinspace s$\\
  \hline
  Online test & $5\thinspace ms$\\
  \hline
  \end{tabular}
  \label{tab:computational_costs}
\end{table}

\begin{figure}[ht]
  \centering
  \includegraphics[width=0.55\columnwidth]{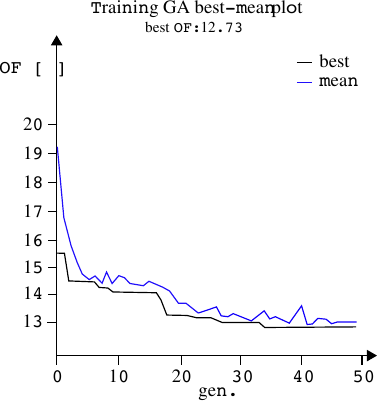}
  \caption{GA best-mean plot.}
  \label{fig:best_mean}
\end{figure}

\noindent The online test computational cost is referred to the single timeslot, i.e. the model takes $5\thinspace ms$ (on average) to decide the optimal power flows for $k+1$. By way of example, the GA best-mean plot for a randomly chosen execution (among the overall $10$) of the model training is presented in Fig.\thinspace\ref{fig:best_mean}.

\noindent Comparisons between the online and the offline approaches about optimal power flows, and $SoE$ are proposed in Tab.\thinspace \ref{tab:comparison}.

\begin{table*}
  \centering
  \begin{tabular}{|c|c|}

    \hline
    \textbf{Offline approach} & \textbf{Online approach} \\
    \hline
  
    \includegraphics[width=0.4\linewidth]{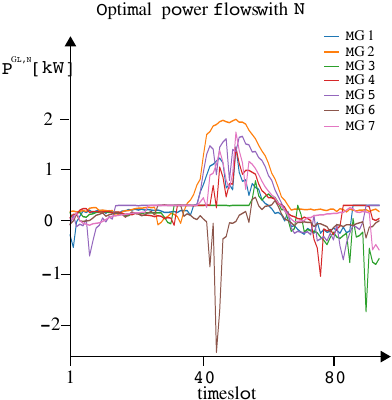} & \includegraphics[width=0.4\linewidth]{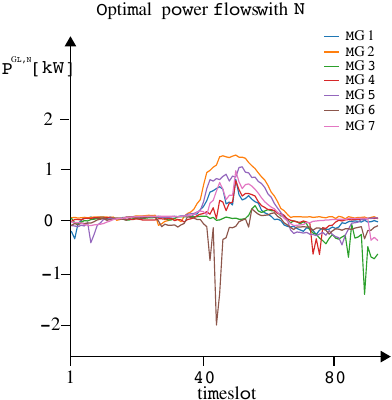} \\
    \hline

    \includegraphics[width=0.4\linewidth]{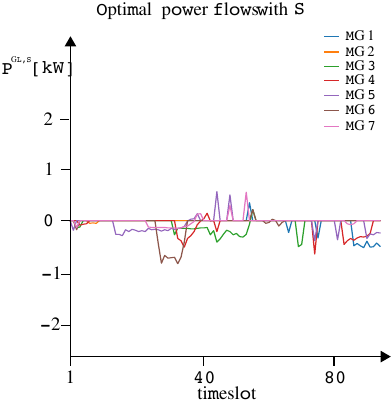} & \includegraphics[width=0.4\linewidth]{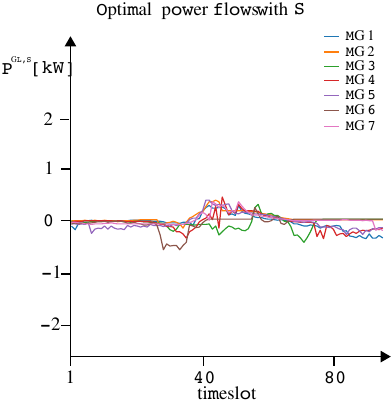} \\
    \hline

    \includegraphics[width=0.33\linewidth]{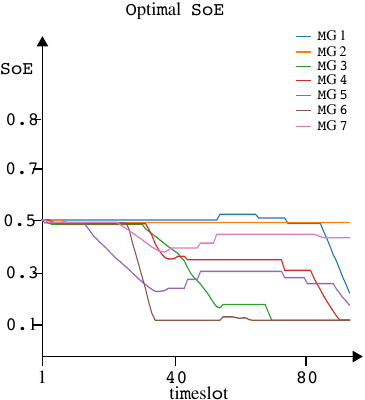} & \includegraphics[width=0.33\linewidth]{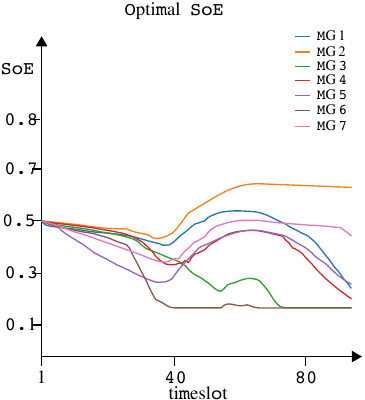} \\
    \hline
    
  \end{tabular}
  \caption{Comparison about optimal power flows and SoE over time between the offline and the online approaches.}
  \label{tab:comparison}
\end{table*}

\begin{figure}[ht]
  \centering
  \includegraphics[width=0.47\columnwidth]{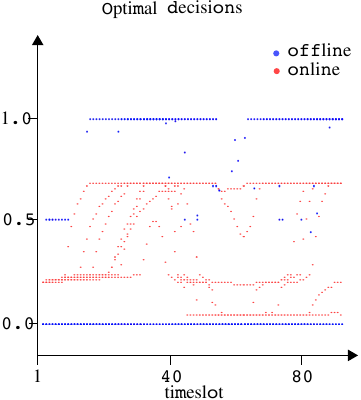}
  \caption{Optimal decisions both for the offline and the online approaches.}
  \label{fig:optimal_alpha}
\end{figure}

\noindent The optimal $\hat{\alpha}$ distribution overt time, both for the offline and the online approaches, are depicted in Fig. \thinspace\ref{fig:optimal_alpha}.

\begin{figure}[ht]
  \centering
  \includegraphics[width=0.47\columnwidth]{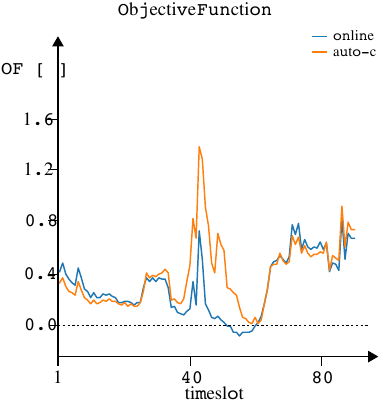}
  \caption{Optimal OF values comparison.}
  \label{fig:fo_vs_auto}
\end{figure}

\noindent In addition, in Fig.\thinspace\ref{fig:fo_vs_auto}, the optimal OF values over time are shown both for the auto-consumption approach and the online approaches, (not for the offline one because an online optimization is the required task for the model).

\subsection{Results discussion}
\label{sub-sec:results_discussion}
\noindent The auto-consumption optimal OF is worst than the benchmark, being the cost higher (Tab.~\ref{tab:hems_train_offline}). It means that a hierarchical optimal flow optimization leads to better results than a local auto-consumption energy management, legitimizing the use of an HEMS for the application at hand. Furthermore, the offline HEMS optimal OF is also better than in the auto-consumption optimization with an error less than $2\thinspace\%$ compared to the benchmark. Therefore, it can be reasonably stated that the HEMS approach is reliable in principle.
\noindent The online HEMS makes and error of about $3\thinspace\%$ compared to the offline procedure and of roughly $5\thinspace\%$ compared to the benchmark (Tab.\thinspace\ref{tab:hems_online}), confirming the model validity. Furthermore, savings of roughly $20\thinspace\%$ compared to the auto-consumption approach are achieved, proving that, like for the offline approach, the online HEMS plenty outperforms auto-consumption.
\noindent Comparing the two approaches by analyzing optimal power flows and $SoE$ values over time, as shown in Tab. \thinspace \ref{tab:comparison}, some further considerations rise. First of all, the most relevant gap occurs at the middle of the day, when the PV production is about to reach its peak. The online HEMS exchanges more power with the Main Grid N than the offline HEMS, both in energy purchasing and in energy selling. As a consequence, power flows involving the batteries are more frequent in the online HEMS and the $SoE$ is varies more compared to the offline one. A reasonable interpretation is that the offline solution is (obviously) sub-optimal because it belongs more to the auto-consumption than in the offline case. That conclusion is strengthen by the optimal decision variables distribution in Fig.\thinspace\ref{fig:optimal_alpha}. Indeed, online HEMS decisions are distributed almost in the middle of the domain while offline HEMS decisions are quite sharply distributed close to the bounds but mainly to the lower one ($0$ represents no auto-consumption). \\
\noindent That said, it could be interesting to know why the HEMS outperforms auto-consumption. By observing Fig.\thinspace\ref{fig:fo_vs_auto}, it can be stated that the OF is better for the HEMS in the middle of the day, when the REC can even get profit (the OF is negative) unlike the auto-consumption case. It could be asserted that, through auto-consumption, the REC is too forced to rely on batteries for satisfying its energy demand, resulting in high wear costs. Rather, a better equilibrium between battery wear costs,  energy purchasing costs, energy selling revenues and incentives is reached by removing the auto-consumption constraint as much as necessary for taking advantage of the PV excess energy in the middle of the day. That conclusion is corroborated by the fact that in Fig.\thinspace\ref{fig:optimal_alpha} the optimal $\hat{\alpha}$ tend to $0$ in the middle of the day.

\subsection{Model Explainability}
\label{sub-sec:model_explainability}
\begin{figure}[ht]
  \centering
  \includegraphics[width=0.47\columnwidth]{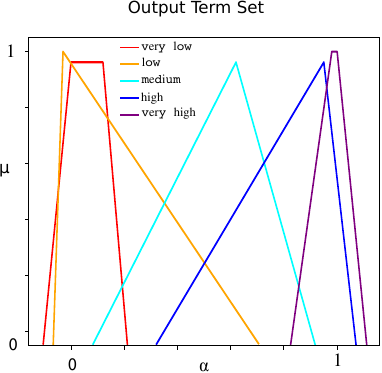}x
  \caption{Optimal output Term Set.}
  \label{fig:output_term}
\end{figure}

\noindent The optimal output Term Set for the online optimization task is shown in Fig. \ref{fig:output_term}

\noindent The optimal Rule Set for the online optimization task is presented below:
\begin{itemize}
    \item If $SoE$ is $Very Low$ then $\alpha$ is $Very High$ $(0.17)$
    \item If $SoE$ is $Low$ then $\alpha$ is $Medium$ $(0.64)$
    \item If $SoE$ is $Medium$ then $\alpha$ is $Medium$ $(0.83)$
    \item If $SoE$ is $High$ then $\alpha$ is $Low$ $(0.83)$
    \item If $SoE$ is $Very High$ then $\alpha$ is $Low$ $(0.22)$
\end{itemize}

\noindent Firs of all, it can be observed that the areas under the MFs in the middle of the output Term Set are larger than the same for MFs at the edge. That is in line with the Rule Set, since the most weighted Rules present $Medium$ and $Low$ Consequents for the output $\alpha$. Such a result could mean that the online HEMS tends to prefer a middle way between complete local self-consumption and the opposite behaviour, as confirmed by Fig. \thinspace \ref{fig:optimal_alpha}. Moreover, the more the $SoE$ is high the more $\alpha$ is low (which implies no self-consumption), reasonably meaning that the HEMS considers to be more convenient to sell excess energy that has an high probability to be a non zero value, since batteries tend to be full of energy from the last timeslot.

\section{Conclusions}
\label{sec:conclusions}
\noindent In thus work, an online HEMS is synthesized for a REC costs minimization, with each participant equipped with a local EMS that maximized its self-consumption while the global decider, aware of the nodes state, overwrites the local optimal decisions about power flows. That design choice comes from the need of exploring more in deep which objective could be more proper for a REC. Thanks to the realism level given by the slavish reference to the EU technical legislation framework, (as inherited by Italy) as well as a proper ESS wear cost model. Since economical details and a REC business model are not included, the final electricity bill costs for each participant are not evaluated. Indeed, a comparison between the self-consumption and the costs minimization approaches is aimed. The XAI FIS-GA optimizer, supported by Neural LSTM PV power predictions, shows that the hierarchical costs minimization strategy outperforms local self-consumption leading to roughly $20 \thinspace \%$ more savings, with a good precision and computational times in the millisecond range. \\

\noindent The proposed algorithm could reach an higher realism level if a REC business model was included in the problem formulation, so that electricity bill savings could be accurately estimated. Since the model is properly parametrized, other Countries technical legislation frameworks would be adopted very easily. To finish, the XAI characterization could lead to a new interesting knowledge about hierarchical strategies like that while new enriched versions are synthesized.

\printbibliography

\end{document}